\documentclass[a4paper,notitlepage,11pt]{article}

\usepackage[latin1]{inputenc}
\usepackage[lmargin=2.5cm, rmargin=2.5cm,bottom=2.5cm,top=2.5cm]{geometry}
\usepackage{amssymb,amsmath,amsfonts,amsthm,amscd,mathtools}
\usepackage{srcltx}
\usepackage{textcomp}
\usepackage{graphics}
\usepackage{color}
\usepackage{pict2e} 

\usepackage{graphicx}
\usepackage{gastex}
\usepackage{pstricks}

\newtheorem{theorem}{Theorem}
\newtheorem{corollary}{Corollary}
\newtheorem{lemma}{Lemma}
\newtheorem{proposition}{Proposition}
\newtheorem{question}{Question}

\theoremstyle{definition}
\newtheorem{definition}{Definition}
\newtheorem{example}{Example}
\newtheorem{remark}{Remark}

\renewcommand{\alph}{\rm{alph}}

\title{Abelian bordered factors and periodicity }

\author{
    \'Emilie Charlier\thanks{Universit\'e de Li\`ege, Belgium, echarlier$@$ulg.ac.be,
        supported in part by the Academy of Finland under FiDiPro grant},
    Tero Harju\thanks{University of Turku, Finland, harju$@$utu.fi},
    Svetlana Puzynina\thanks{Corresponding author, LIP, ENS de Lyon, France,
        and Sobolev Institute of Mathematics, Russia,
        s.puzynina$@$gmail.com}
    and Luca Q. Zamboni\thanks{Universit\'e de Lyon 1, France, zamboni$@$math.univ-lyon1.fr,
        supported in part by the Academy of Finland under FiDiPro grant}
}

\date{}

\begin{document}
\maketitle
\thispagestyle{empty} 
\pagestyle{empty}

\begin{abstract} 
    A finite word $u$ is said to be \emph{bordered} if $u$ has a proper prefix
which is also a suffix of $u$, and \emph{unbordered} otherwise.
    Ehrenfeucht and  Silberger 
    proved that an infinite word is purely periodic
    if and only if it contains only finitely many unbordered factors.
    We are interested in abelian and weak abelian analogues of this result;
 namely, we investigate the following question(s):
    Let $w$ be an infinite word such that  all sufficiently long factors are (weakly) abelian bordered;
    is $w$ (weakly) abelian periodic? 
    In the process we answer a question of  
    Avgustinovich et al.
    concerning the abelian critical factorization theorem.
    \end{abstract}

\section{Introduction}
A finite word $u$ is \emph{bordered}, if there exists a non-empty
word $z$ which is a proper prefix and a suffix of $u$. The
following theorem states a well-known connection between
periodicity and unbordered factors:

\begin{theorem}\emph{\cite{EhrSil,Saari}} \label{Theorem1}
    An infinite word $w$ is purely periodic if and only if
    there exists a constant $C$ such that every factor $v$ of $w$ with $|v|\geq C$ is bordered.
\end{theorem}

In this paper we are interested in extending this result to
the abelian and weak abelian settings.

Two finite words $u$ and $v$ are \emph{abelian equivalent} if and only if for
each letter $a,$ the number of occurrences of $a$ in $u $ (denoted $|u|_a)$ is equal to $|v|_a$. 
In other words, $u$ and $v$ are permutations of one another. 
An infinite word $w$ is called \emph{abelian ultimately periodic} if $w = u v_1v_2v_3 \cdots$, where
the $v_i$ are pairwise abelian equivalent.

    An infinite word $w$ is called \emph{weakly abelian ultimately periodic} 
    if $w = u v_1v_2v_3 \cdots$, where $v_i$ are finite words with the same frequencies of letters for $i\geq 1$.
In other words, a word is weakly abelian  ultimately periodic if it can be
factored into some prefix $u$ followed by infinitely many words of
possibly different lengths with the same letter frequencies. For
more on weak abelian periodicity see \cite{wap}.

We say that a finite
word $u$ is \emph{abelian bordered} if $u$ contains a non-empty proper
prefix which is abelian equivalent to a suffix of $u.$ 
Abelian bordered words were recently considered in
\cite{CCCI,RRS}. A finite word $u$ is \emph{weakly abelian
bordered} if it has a non-empty proper prefix and a suffix with
the same letter frequencies.

In Sections \ref{wab2} and \ref{wabk}, we consider weak abelian
borders in the binary and non-binary cases, respectively. We prove
that the condition on finitely many weak abelian unbordered
factors implies weak abelian periodicity plus some additional
restrictions (Theorems \ref{wap} and \ref{wap1}). 
In Section~\ref{ab}, we consider infinite words having only finitely many abelian unbordered factors. 
We do not know if such words are necessarily abelian periodic. 
However, we are able to show that such words have bounded abelian complexity.
In Section~\ref{5}, we provide an answer to a question from
\cite{akp} concerning abelian critical factorization theorem. 
In the last section we summarize the results and propose some open problems.

\section{Preliminaries}

In this section we give some basics on words following terminology
from \cite {Lo} and introduce our notions.

Given a finite non-empty set $\Sigma$ (called the alphabet), we
denote by $\Sigma^*$ and $\Sigma^{\omega}$, respectively, the set
of finite words and the set of (right) infinite words over the
alphabet $\Sigma$. Given a finite word $u = u_1 u_2 \cdots u_n$
with $n \geq 1$ and $u_i \in \Sigma$, we denote the length $n$ of $u$ by $|u|$.
The empty word will be denoted by $\varepsilon$ and we set $|\varepsilon| = 0$.
Given the words $w$, $x$, $y$, $z$ such that $w = xyz$, $x$ is called a \emph{prefix}, 
$y$ is a \emph{factor} and $z$ a \emph{suffix} of $w$.
The word $x$ is a \emph{proper prefix} if $0<|x|<|w|$.
We let $w[i, j] = w_iw_{i+1} \cdots w_j$ denote the factor starting at position $i$ and ending at position $j$
of $w=w_1w_2w_3\cdots$, where the $w_k\in\Sigma$.
An infinite word $w$ is \emph{ultimately periodic} (or briefly \emph{periodic})
if for some finite words $u$ and $v$ it holds $w=uv^{\omega} = uvv\cdots$; $w$ is
\emph{purely periodic} if $u=\varepsilon$.
An infinite word is \emph{aperiodic} if it is not ultimately periodic.
Given a finite word $u = u_1 u_2 \cdots u_n$ with $n \geq 1$ and
$u_i \in \Sigma$, for each $a \in \Sigma$, we let $|u|_a$ denote
the number of occurrences of the letter $a$ in $u$.
Two words $u$ and $v$ in $\Sigma^*$ are \emph{abelian equivalent},
denoted $u\sim_{ab} v$, if and only if $|u|_a = |v|_a$ for all $a\in \Sigma$.
It is easy to see that abelian equivalence is an equivalence relation on $\Sigma^*$.
An infinite word $w$ is called \emph{abelian ultimately periodic} 
(or briefly \emph{abelian periodic})
if $w = u v_1v_2v_3 \cdots$, where  $v_i \sim_{ab} v_j$ for all integers $i, j \geq 1$.

For a finite non-empty word $w\in\Sigma^*$, we define the \emph{frequency}
$\rho_a(w)$ of a letter $a\in\Sigma$ in $w$ as $\rho_a(w)=\frac{|w|_a}{|w|}$.
For an infinite word $w\in\Sigma^\omega$ and a letter $a\in\Sigma$, if the limit
\[
\lim_{n\to\infty}\rho_a(w[1..n])
\]
exists, then we define the \emph{frequency} $\rho_a(w)$ of $a$ in
$w$ to be the latter limit.
The infinite word $w$ has
\emph{uniform letter frequencies} if, for every letter $a$ of $w$, the
ratio $\frac{|w[k..k+n]|_a}{ n+1 }$ has a limit $\rho_a(w)$ when $n\to \infty$, uniformly in $k$.
In general the existence of uniform
frequencies is stronger than (prefix) frequencies, but in our
consideration in fact all the words have uniform frequencies.

Now we provide some background on weak abelian periodicity from
\cite{wap}.

\begin{definition}
    An infinite word $w$ over an alphabet $\Sigma$ is called \emph{weakly abelian ultimately periodic} (or briefly \emph{weakly abelian periodic})
    if $w = u v_1v_2v_3 \cdots$, where  $\rho_a(v_i) = \rho_a(v_j)$
    for all $a\in\Sigma$ and all integers $i, j\geq 1$.
\end{definition}

An infinite word $w$ is called \emph{bounded weakly abelian
periodic}, if it is weakly abelian periodic with bounded lengths of
blocks, i.e., there exists $C$ such that for every $i$ we have $|v_i|\leq C$.

\begin{example}
    The word $(01)(0^21^2)(0^31^3)(0^41^4)\cdots$ is weakly abelian periodic
    but not bounded weakly abelian periodic.
\end{example}

 We make use of the following geometric interpretation
of weak abelian periodicity.
We translate an infinite word $w= w_1 w_2 \cdots \in
\Sigma^{\omega}$ to a graph visiting points of the lattice
$\mathbb{Z}^{|\Sigma|}$ by interpreting letters of $w$ as drawing
instructions. In the binary case we associate $0$
with a move by the vector $\textbf{v}_0=(1,-1)$, and
$1$ with a move by $\textbf{v}_1=(1,1)$.
We start at the origin $(x_0,y_0)=(0,0)$. 
At step $n$, we are at a point $(x_{n-1},y_{n-1})$
and we move by a vector corresponding to the letter $w_{n}$, so
that we come to a point $(x_{n},
y_{n})=(x_{n-1},y_{n-1})+\textbf{v}_{w_n}$,
and the two points $(x_{n-1}, y_{n-1})$ and $(x_{n}, y_{n})$
are connected with a line segment.
We let $G_w$ denote the
corresponding graph, and $G_w(n)=(x_n,y_n)$.

\begin{lemma}
A weakly abelian periodic word $w$ has a graph with infinitely many integer points
on some line with a rational slope.
\end{lemma}


We remark that instead of the vectors $(1,-1)$ and $(1,1)$,
one could use any other pair of noncollinear vectors $\textbf{v}_0$
and $\textbf{v}_1$.
For a $k$-letter alphabet one can consider a similar graph in
$\mathbb{Z}^k$ with linearly independent integer vectors
$\textbf{v}_1, \dots, \textbf{v}_k$.
For example, one can take as $\textbf{v}_i$ a unit vector with $1$ at position $i$
and all other coordinates equal to $0$.

 In the binary case and for vectors
$\textbf{v}_0=(1,-1)$ and $\textbf{v}_1=(1,1)$, it will be
convenient for us to use a function $ g_w\colon n\mapsto y_n $, so
that $G_w(n) = (n, g_w(n))$.
It can also be regarded as a piecewise linear function with line
segments connecting integer points (see an example on Figure 1).

\begin{figure}[h]
    \begin{center}   \unitlength=1pt
        \begin{picture}(330,60)
        \multiput(10,0)(10,0){32}{\line(0,1){60}}
        \multiput(0,10)(0,10){5}{\line(1,0){330}}

        \thicklines \put(0,30){\vector(1,0){330}}
        \put(10,0){\vector(0,1){60}} \put(10,30){\circle*{5}}

        \put(10,30){\line(1,-1){10}} \put(20,20){\line(1,1){20}}
        \put(40,40){\line(1,-1){10}} \put(50,30){\line(1,1){10}}
        \put(60,40){\line(1,-1){20}} \put(80,20){\line(1,1){20}}
        \put(100,40){\line(1,-1){20}} \put(120,20){\line(1,1){10}}
        \put(130,30){\line(1,-1){10}} \put(140,20){\line(1,1){20}}
        \put(160,40){\line(1,-1){10}} \put(170,30){\line(1,1){10}}
        \put(180,40){\line(1,-1){20}} \put(200,20){\line(1,1){10}}
        \put(210,30){\line(1,-1){10}} \put(220,20){\line(1,1){20}}
        \put(240,40){\line(1,-1){20}} \put(260,20){\line(1,1){20}}
        \put(280,40){\line(1,-1){20}} \put(300,20){\line(1,1){10}}
        \put(310,30){\line(1,-1){10}}
        \end{picture}
            \caption{The graph of the Thue-Morse word with $\textbf{v}_0=(1,-1)$,
                    $\textbf{v}_1=(1,1)$. }\label{table}
    \end{center}
\end{figure}
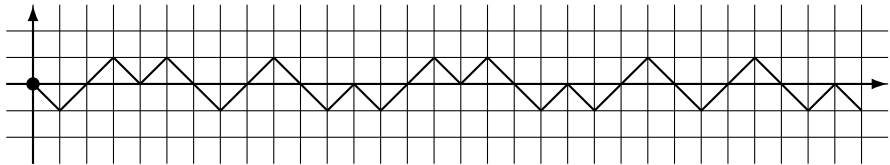

\section{Weak abelian borders: binary alphabet}\label{wab2}

We proceed with the following theorem relating weak abelian
borders and weak abelian periodicity, giving one way analog of the
characterization of periodicity in terms of unbordered factors (Theorem~\ref{Theorem1}):

\begin{theorem} \label{wap}
    Let $w$ be an infinite binary word.
    If there exists a constant $C$ such that every factor $v$ of $w$ with $|v|\geq C$
    is weakly abelian bordered, then $w$ is bounded weakly abelian periodic.
    Moreover, its graph lies between two rational lines
    and has points on each of these two lines with bounded gaps.
\end{theorem}

Remark that following the terminology from \cite{wap},
 a bounded weakly abelian periodic word is necessarily of bounded width.
Namely, a word $w$  is of \emph{bounded width} if there exist two lines with the same slope,
so that the graph of $w$ lies between these two lines.
Formally, there exist numbers $\alpha, \beta, \beta'$
so that $\alpha x+\beta \leq g_w(x)\leq \alpha x+\beta'$.
We note that in the case of weak abelian periodicity $\alpha$ must be rational.

We will also need the notion of balance.
A word $w\in\Sigma^{\omega}$ is called $K$\emph{-balanced}, if for each
letter $a$ and two factors $u, v$ of $w$ such that $|u|=|v|$ the
inequality $||u|_a-|v|_a|\leq K$ holds.
We simply say that w is \emph{balanced} if it is $K$-balanced for some $K$.
We remark that balance implies the existence of letter frequencies:

\begin{proposition} \cite[Proposition~2.4]{Berthe-Delecroix}\label{24}
An infinite word $w \in \Sigma^{\omega}$ is balanced if and
only if it has uniform letter frequencies and there exists a
constant $B$ such that for any factor $u$ of $w$ and for all letters $a\in
\Sigma$, we have $\big| |u|_a - \rho_a (w) |u|\big| \leq B$.
\end{proposition}

In fact, the latter proposition implies that balance is equivalent to
the bounded width property.
Indeed, the bounded width property implies balance.
The converse is obtained by applying the
above proposition to the prefixes of the word.

\begin{corollary}
An infinite word is balanced if and only if it is of bounded width.
\end{corollary}

The proof of Theorem \ref{wap} heavily relies on the graph representation of the word.
It consists of several lemmas restricting the form of the word.
We first prove that if a factor $w[i+1..j]$
of $w$ is such that the graph of the word between
$i$ and $j$ lies above or below the line segment connecting the
points $(i, g_w(i))$ and $(j, g_w(j))$, then $w[i+1..j]$ is weakly abelian unbordered:

\begin{lemma}\label{wab_factor}
    Let $w$ be an infinite binary word, and $i$, $j$ be integers, $i<j$.
    If $g_w(k) >  \frac{g_w(j) - g_w(i) }{j-i}k
    +  \frac{g_w(i)j - g_w(j) i}{j-i}$ for each $ i< k< j$,
    then the factor $w[i+1..j]$ is weakly abelian unbordered.
\end{lemma}

We also note that a symmetric
assertion holds for the case when the graph lies below this line:
simply invert the inequality in the statement of the lemma.

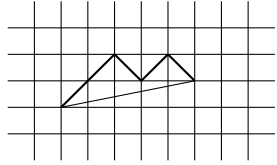
\begin{figure}[h]
    \begin{center}
    \unitlength=1pt
        \begin{picture}(100,60)
        \multiput(10,0)(10,0){9}{\line(0,1){60}}
        \multiput(0,10)(0,10){5}{\line(1,0){100}}
        \thicklines 
        \put(20,20){\line(1,1){20}}
        \put(40,40){\line(1,-1){10}}
        \put(50,30){\line(1,1){10}}
        \put(60,40){\line(1,-1){10}}
	\thinlines
	\put(20,20){\line(5,1){50}}
        \end{picture}
      \caption{Illustration of Lemma \ref{wab_factor}:
            weakly abelian unbordered factor  11010.}
      \label{fig_lemma1}
      \end{center}
\end{figure}

\begin{proof}
     For any proper prefix $p$ and any proper suffix $s$ of $w[i+1..j]$
     we have $\rho_1(p)-\rho_0(p)> \frac{g_w(j) - g_w(i)}{j-i}$
     and $\rho_1(s)-\rho_0(s)< \frac{g_w(j) - g_w(i)}{j-i}$.
     Hence there is no weak abelian border.
\end{proof}

\begin{figure}[htbp]
    \begin{center}
    \unitlength=1pt
        \begin{picture}(160,90)
        \multiput(10,0)(10,0){15}{\line(0,1){90}}
        \multiput(0,10)(0,10){8}{\line(1,0){160}}
        \thicklines
        \put(20,40){\line(1,1){30}}
        \put(50,70){\line(1,-1){50}}
    \put(100,20){\line(1,1){20}}
        \put(120,40){\line(1,-1){30}}\thinlines
        \put(20,40){\line(13,-3){130}}
            \end{picture}
      \caption{The reciprocal of Lemma \ref{wab_factor} is not true:
            weakly abelian unbordered factor  $1110000011000$.}
      \label{counterex}
      \end{center}
\end{figure}
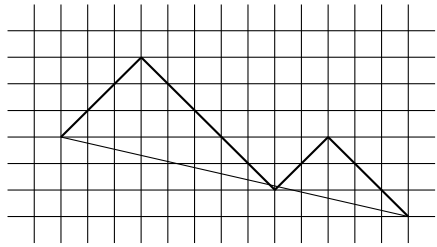

However, the reciprocal of the previous lemma does not hold.
A counterexample is given by the weakly abelian unbordered word $1110000011000$:
see Figure~\ref{counterex}.

\begin{lemma}\label{bal}
     Let $w$ be an infinite binary word with finitely many weakly abelian unbordered factors.
Then $w$ is balanced.
\end{lemma}

We remind that by balance we mean $K$-balance for some $K$.

\begin{proof}
    Suppose the converse.
    Unbalance implies that for every $K$ there exist $i$, $j$, $l$ with $i<l<j$, such that
    \begin{equation}\label{eq_bal}
            \left|g_w(l) - \frac{g_w(j) - g_w(i) }{j-i} l
                - \frac{g_w(i)j - g_w(j) i}{j-i} \right| \geq K.
    \end{equation}
    In other words, the point $(l, g_w(l))$ lies at distance at least $K$
    from the line connecting the points $i$ and $j$ (distance measured vertically).
    Indeed, unbalance implies that there exist two factors $u$ and $v$ of the same length
    in which the number of occurrences of the letter $0$ (and hence also the letter $1$) differs by more than $K$.
    Consider the factor $w[i+1..j]$ starting with $u$ and ending with $v$. 
	It is not hard to see that it satisfies the inequality
    \eqref{eq_bal} for  either $l=i+|u|$ or $l=j-|v|$.

    Take the shortest such factor (i.e., choose $i$, $j$ satisfying
    the condition \eqref{eq_bal} with the smallest $j-i$).
    Then the graph $G_w$ does not intersect the line segment
    connecting the points $(i, g_w(i))$ and $(j, g_w(j))$,
    for otherwise there exists a shorter factor satisfying \eqref{eq_bal}.
    Hence due to Lemma \ref{wab_factor}, it remains to notice that the lengths
    of such factors grow as $K$ grows (in fact, $j-i>K$).
\end{proof}

Since balance implies the existence of letter frequencies,
Lemma \ref{bal} implies that a word satisfying the conditions of
Theorem \ref{wap} has letter frequencies. We will now prove
that the frequencies are rational.


\begin{lemma}\label{freq}
     Let $w$ be an infinite binary word with finitely many weakly abelian unbordered factors.
Then the letter frequencies are rational.
\end{lemma}

\begin{proof}
Balance implies bounded width, i.e.,
there exist $\alpha,$ $\beta',$ $\beta''$ such that the graph of
$w$ lies between two lines: $ \alpha x+\beta'\leq g_w(x)\leq \alpha x+\beta''$.
 Here $\alpha$ is related to the letter frequencies as follows:
 $\rho_0=\frac{1-\alpha}{2}$ and
$\rho_1=\frac{1+\alpha}{2}$.
So, we need to prove that $\alpha$ is rational.
Assume the converse.
Take the lowest (in the sense of the smallest $\beta$) line $y=\alpha x + \beta$ such that there
are points of the graph arbitrarily close to it.
We build the line (and prove its existence) as follows.

Take $\varepsilon>0$, divide the stripe $ \alpha x+\beta'\leq
y\leq \alpha x+\beta''$ into stripes of width at most
$\varepsilon$ (width can be measured vertically, although this is not important).
At least one of the stripes contains infinitely
many points $(i, g_w(i))$ of the graph $G_w$.
Choose the lowest such stripe $ \alpha x+\beta'_1\leq y \leq \alpha x+\beta''_1$.
Divide it into stripes of width $\varepsilon_1< \varepsilon$,
continue the process with $\varepsilon_n\to 0$.
The value of $\beta$ in the desired line is obtained as a limit
$\beta=\lim_{n\to\infty}\beta'_n=\lim_{n\to\infty}\beta''_n$.
To prove that $\alpha$ is rational, we consider three cases:

\textbf{Case 1:}
There are no points of the graph $G_w$ below the line $y=\alpha x + \beta$.
In this case take $\varepsilon>0$ and consider the stripe
$\alpha x+\beta\leq y\leq \alpha x+\beta+\varepsilon$.
There are infinitely many points in this
stripe by the definition of the line $y=\alpha x + \beta$.
Consider any two consecutive points $(n_1, g_w(n_1))$ and $(n_2,
g_w(n_2))$ from the stripe (consecutive in the sense that for any
$n$, $n_1<n<n_2$, we have $g_w(n)>\alpha x+\beta+\varepsilon$).
By Lemma \ref{wab_factor}, the factor $w[n_1+1..n_2]$ is weakly abelian
unbordered, since the points of the graph $G_w$ between $n_1$ and
$n_2$ lie above the line $y= \alpha x+\beta+\varepsilon$, which in
turn lies above the line segment connecting the points $(n_1,
g_w(n_1))$ and $(n_2, g_w(n_2))$.

Now we will prove that the minimal possible length $(n_2-n_1)$
grows as $\varepsilon\to 0$. Indeed, take the shortest factor
$w[n_1+1..n_2]$ with $n_1$, $n_2$ consecutive points in the stripe
such that the slope of the line segment connecting the points
$(n_1, g_w(n_1))$ and $(n_2, g_w(n_2))$ is the closest to
$\alpha$, i.e., $\big|\alpha - |\frac{g_w(n_2)-g_w(n_1)}{n_2-n_1}|\big|$
is minimal among the factors of the length $(n_2-n_1)$ with both
ends in the stripe.
If there are several such factors, choose one
of them, say the leftmost.
Since $\alpha$ is irrational, it is not equal to the slope of the line connecting points
$(n_1,g_w(n_1))$ and $(n_2, g_w(n_2))$.
So, take $\varepsilon'<\big||g_w(n_2)-g_w(n_1)|-\alpha(n_2-n_1)\big|$,
then no two points of the graph corresponding to positions in the word at
distance $(n_2-n_1)$ can simultaneously be in the stripe $ \alpha
x+\beta\leq y\leq \alpha x+\beta+\varepsilon'$.
So, the minimal distance between points in the new (thinner) stripe of width
$\varepsilon'$ increased, which corresponds to weak abelian
unbordered factors of length more than $n_2-n_1$.
Continuing this line of reasoning
and taking smaller and smaller $\varepsilon$, we get arbitrary
long weakly abelian unbordered factors.

\textbf{Case 2:}  There are finitely many points of the graph
$G_w$ below the line $y=\alpha x + \beta$ (but the number of
such points is nonzero). In this case take the last point $(n_0,
g_w(n_0))$ below the line: $g_w(n_0)\leq\alpha n_0 + \beta$, and
$g_w(n)>\alpha n + \beta$ for each $n>n_0$. Consider a prefix of
length $N>n_0$ of $w$, and the part of the graph between $n_0$ and
$N$. Choose a point $n'$, $n_0<n'\leq N$, such that the point
$(n', g_w(n'))$ is the closest to the line among the points $(n,
g_w(n))$, $n_0<n\leq N$:
$$|g_w(n')-\alpha n' - \beta| = \min_{n_0<n\leq N} |g_w(n)-\alpha n -
\beta|.$$
 Then due to Lemma
\ref{wab_factor}, the factor $w[n_0+1..n']$ is weak abelian
unbordered. Indeed, the points of the graph $G_w$ between $n_0$
and $n'$ lie above the line $y= \alpha x+g_w(n')-\alpha n'$ (the
line parallel to $y=\alpha x + \beta$ and passing through the
point $(n', g_w(n'))$), which in turn lies above the line segment
connecting the points $(n_0, g_w(n_0))$ and $(n', g_w(n'))$:
\[
g_w(n)\geq\alpha n+g_w(n')-\alpha n'
    > \frac{g_w(n') -g_w(n_0) }{n'-n_0}n
        + \frac{g_w(n_0)n' - g_w(n') n_0}{n'-n_0}
\]
for $n_0<n<n'$. Since there are points of the
graph arbitrary close to the line $y=\alpha x + \beta$, then,
increasing $N$, we will find larger and larger values $n'$ and
hence find arbitrary long weakly abelian unbordered factors.

\textbf{Case 3:} There are infinitely many points of the graph
$G_w$ below the line $y=\alpha x + \beta$.
Since we chose the lowest line with the property that there are infinitely many
points arbitrary close to the line, it follows that for any
$\varepsilon>0$ there are only finitely many points below the line
$y=\alpha x + \beta-\varepsilon$ (if any).
Take $\varepsilon$ such that there are points below the line $y=\alpha x +
\beta-\varepsilon$.
By the condition of Case 3 there are infinitely many points
in the stripe $\alpha x + \beta -\varepsilon \leq y \leq \alpha x + \beta$.
For each two consecutive points in the stripe the corresponding factor is weakly
abelian unbordered due to Lemma \ref{wab_factor}.
As $\varepsilon\to 0$, the lengths of such factors grow for
irrational $\alpha$ (the proof of this is similar to the proof of
the similar fact from Case 1).

Therefore, we came to a contradiction in each case. So the
frequency must be rational.
\end{proof}

\begin{proof}[Proof of Theorem~\ref{wap}.]
Lemmas \ref{bal} and \ref{freq} imply that the graph of the word lies between two lines
of rational slope $\alpha$: $\alpha n + \beta'\leq
g_w(n)\leq \alpha n + \beta''$ for each $n$.
 We can choose such lines
with the minimal $\beta''$ and maximal $\beta'$.
To prove the theorem it remains to show that the graph of the word has points
on each of these two lines with bounded gaps.
We provide a proof for the line $y=\alpha x + \beta'$, for the other line the proof is symmetric.

First we prove that there are points of the graph on the line.
Clearly, for rational $\alpha$ there is a fixed minimal distance
from the integer points of the lattice $\mathbb{Z}^2$ to the line
(note that for irrational slope this does not hold). If there are
no points of the graph on the line, since the slope is rational,
instead of $y=\alpha x + \beta'$ we could choose a different line
$y=\alpha x + \beta'''$ with the same slope passing through the
point of the graph closest to the line, and with
$\beta'''>\beta'$, which contradicts the choice of $\beta'$ as the
maximal possible.

Secondly, there are infinitely many points on the line.
Otherwise 
consider a line $y=\alpha x + \beta'''$ with $\beta'''>\beta'$, with
infinitely many points of the graph on the line and with the
smallest such $\beta'''$.
Note that here we need again $\alpha$ to be rational.
Then there exists $n_0$ such that for all $n>n_0$ we
have $g_w(n)\geq \alpha n + \beta'''$, i.e., starting from $n_0$
the graph lies above the line.
Taking the smallest possible $n_0$
satisfying the condition, we get that the point $(n_0, g_w(n_0))$
lies below the line.
Hence for each $n'>n_0$ corresponding to a
point on the line, i.e., with $g_w(n')=\alpha n' + \beta'''$, we
have that the factor $w[n_0+1..n']$ is weakly abelian unbordered by
Lemma \ref{wab_factor}.
Indeed, all the points between $n_0$ and
$n''$ lie above the line $y=\alpha x + \beta'''$ and hence above
the line segment connecting the points $(n_0,g_w(n_0))$ and
$(n',g_w(n'))$ of the graph corresponding to the beginning and the
end of the factor.

Finally, we need to prove that the points on the line $y=\alpha x
+ \beta'$ appear in bounded gaps.
This follows from the fact that for two consecutive points $n_1,$ $n_2$ on the line $y=\alpha x +
\beta'$ the factor $w[n_1+1..n_2]$ is weakly abelian unbordered by
Lemma \ref{wab_factor}.
\end{proof}

\begin{remark} Except for the case $\rho_0=\rho_1=1/2$
(equal frequencies), we do not know whether the converse of
Theorem 2 is true (see Question~\ref{question2}).
In the case of equal
frequencies the converse holds, and it could be easily seen from
the graph of the word.
In fact, we will prove that all
sufficiently long factors have weak abelian borders with
frequencies $1/2$.
Frequencies equal to $1/2$ correspond to
$\alpha=0$, so the graph lies between two horizontal lines
$y=\beta'$ and $y=\beta''$, and intersects each of them with
bounded gaps.
Denote the maximal gap by $C$.
Consider a factor $w[i+1..j]$ of length $>C$: $j-i> C$. 
Then there exist
$n',n''$, $i<n',n''<j$, such that $g_w(n')=\beta'$,
$g_w(n'')=\beta''$. Without loss of generality $n'<n''$. For each
integer $\beta$, $\beta'\leq\beta\leq\beta''$, there exists $n$,
$n'\leq n \leq n''$, such that $g_w (n)=\beta$.
In other words, in
the interval $[n', n'']$ the graph passes through any integer
horizontal line between $\beta'$ and $\beta''$. In particular,
there exist $i'$, $j'$, $i<n'\leq i',j'\leq n''<j$, such that
$g_w(i)=g_w(i')$, $g_w(j)=g_w(j')$. 
So, the prefix $w[i+1..i']$ and the suffix $w[j'+1..j]$ 
give a weak abelian border with the frequencies $1/2$. 
But we do not see how to generalize this
observation to the case of nonequal frequencies, since in general
the graph can ``jump'' through several rational non-horizontal
lines and it does not have to pass through all rational lines
between points on lower and upper lines.
\end{remark}

\section{Weak abelian borders: nonbinary alphabet}\label{wabk}

Now we continue with non-binary alphabets and prove the analogue of
Theorem \ref{wap}. 
Basically, a non-binary word with finitely many
weakly abelian unbordered factors must also be balanced, weakly
abelian periodic and have rational letter frequencies. 
To state the analogue of the additional condition in Theorem \ref{wap}
(graph lying between two parallel rational lines and bounded weak
abelian periodicity along those lines) we need some notation.

First, in the non-binary case, instead of bounded width we define cylinder set. 
Let $\emph{\textbf{l}}$ denote a line in $\mathbb{R}^k$,  
$\emph{\textbf{l}}\colon x_i=x_i^0+a_i t$, $i=1,
\dots, k$, and let $M$ be a constant.
A $k$-dimensional \emph{cylinder} $C (\emph{\textbf{l}},M)$ 
is defined as
\[
	C (\emph{\textbf{l}},M) 
		= \{ \bar{x} \in \mathbb{R}^k | d(\emph{\textbf{l}}, \bar{x})\leq M\}
\]
where $d(\emph{\textbf{l}}, \bar{x})$ denotes the distance between the
point $\bar{x}$ and the line $\emph{\textbf{l}}$. 
The line $\emph{\textbf{l}}$ is the \emph{axis} of the cylinder.

Now we will give an analogue of the two parallel lines such that the
graph lies between those two lines and has points on each of those
lines with bounded gaps. 
If the graph $G_w$ of an infinite word
$w$ over a $k$-letter alphabet belongs to some cylinder $C
(\emph{\textbf{l}},M)$, we define a \emph{tangential hyperplane}
to the graph of $G_w$ as a hyperplane $H\colon b_1 x_1 + \dots + b_k x_k = d$ parallel to
the line $\emph{\textbf{l}}$ such that for each point
$\bar{x}=(x_1, \dots, x_k)\in G_w$ one has $b_1 x_1 + \dots + b_k
x_k \geq d$, and in addition there is a point $\bar{x}'=(x_1',\ldots,x_n')\in G_w$
such that $b_1 x'_1 + \dots + b_k x'_k = d$. 
We remark that a tangential hyperplane to the graph $G_w$ does not have to be a
``proper'' tangential hyperplane to the cylinder $C
(\emph{\textbf{l}},M)$, but it is a hyperplane parallel to it. 
In the remaining part of the text, by ``tangential'' we always mean tangential to
the graph. Next, we define a \emph{tangential line}
$\emph{\textbf{l}}'$ to $G_w$, $\emph{\textbf{l}}'\colon x_i=x_i'^0+a_i t$, $i=1, \dots, k$, as a line which belongs to
some tangential hyperplane $H$ to $G_w$, $H\colon b_1 x_1 + \dots +
b_k x_k = d$, such that for each point $\bar{x}\in G_w \cap H$ one
has also $\bar{x}\in \emph{\textbf{l}}'$. 
In other words, a tangential line $\emph{\textbf{l}}'$ is defined as a line parallel
to the cylinder lying in some tangential hyperplane, such that the
only points of the graph belonging to the hyperplane also belong to the line. 
We remark that this line can be inside the cylinder.

Now we are ready to state the non-binary analogue of Theorem
\ref{wap}:

\begin{theorem} \label{wap1}
    Let $w$ be an infinite word over a $k$-letter alphabet. 
    If there exists a constant $C$ such that every factor $v$ of $w$ with $|v|\geq C$
    is weakly abelian bordered, then $w$ is bounded weakly abelian periodic.
    Moreover, its graph $G_w$ belongs to a $k$-dimensional cylinder with axis with rational coefficients,
    and each tangential line to $G_w$ has points of $G_w$ on it with bounded gaps.
\end{theorem}

We emphasize that in fact Theorem \ref{wap} is a particular case of Theorem \ref{wap1}. 
We state and prove it separately for several reasons. 
First, for clarity reasons: the statement of Theorem \ref{wap} is easier and more intuitive. 
Second, key lemmas in non-binary case are simple consequences of those in the binary case. 
On the other hand, the proof of the body of the theorem is more involved, 
so there is not much repetition in the proofs.

First we prove that Lemmas \ref{bal} and \ref{freq} also hold for
the non-binary case:

\begin{lemma}\label{bal1}
       If an infinite word 
   has finitely many weakly abelian unbordered factors,
    then it is balanced.
\end{lemma}

\begin{proof}
	Assume that $w$ is an infinite word 
    with finitely many weakly abelian unbordered factors.
    For $a\in\alph(w)$, consider a morphism $h_a\colon a\mapsto a,\
    b\mapsto c$ for all $b\neq a$ where $c$ is a letter not belonging to $\alph(w)$.
    We call such a morphism a \emph{projection}.
    Clearly, if two words $u$ and $v$ have the same letter frequencies, 
	then so do $h_a(u)$ 
    and $h_a(v)$, for each $a\in \alph(w)$. 
    Hence if $u$ is weakly abelian bordered, then so is $h_a(u)$.
Then $h_a(w)$ has finitely many weakly abelian unbordered factors.
    Now, since a projection of $w$ is a binary word (over the
    alphabet $\{a,c\}$), we can apply Lemma \ref{bal}, hence $h_a(w)$
    is also balanced:
    There exists a constant $K_a$, such that for
    each two factors $u, v$ of $h_a(w)$ such that $|u|=|v|$ the
    inequality $||u|_a-|v|_a|\leq K_a$ holds.
    Taking $K=\max_{a\in \Sigma} K_a$, we get that $w$ itself is $K$-balanced by definition.
\end{proof}

\begin{lemma}\label{freq1}
    If an infinite word has finitely many weakly abelian unbordered factors,
    then its letter frequencies are rational.
\end{lemma}

\begin{proof}
    Clearly, the frequency of any letter $a$ exists in an
    infinite word $x$ if and only if the frequency of the letter $a$
    exists in its projection $h_a(x)$, and these frequencies are equal.
    Since the projection is a binary word, Lemma \ref{freq}
    implies that $w$ has letter frequencies and they are rational.
\end{proof}

We let $(\bar{b}, \bar{x})$ denote the scalar product of vectors
$\bar{b}$ and $\bar{x}\colon (\bar{b}, \bar{x}) = b_1 x_1 + \dots + b_k x_k $.
We make use of the following analogue of Lemma~\ref{wab_factor}.

\begin{lemma}\label{wab_factor1}
    Let $w$ be an infinite word, and $i$, $j$ be integers, $i<j$.
    If there exists a hyperplane $H\colon b_1 x_1 + \dots + b_k x_k = d$ such that
    $G_w(i),  G_w(j) \in H$ and  $(\bar{b},
    G_w(l))>d$ for each $ i< l< j$ (or $(\bar{b}, G_w(l))<d$ for each $ i< l< j$),
    then the factor $w[i+1..j]$ is weakly abelian unbordered.
\end{lemma}

\begin{proof}
    Two factors $w[i_1+1..j_1]$ and $w[i_2+1..j_2]$ 
	share the same letter frequencies if the
    vectors $G_w(j_1)-G_w(i_1)$ and $G_w(j_2)-G_w(i_2)$ connecting
    their ends are collinear.
    Moreover, they differ by a positive multiple.
    For any proper prefix $w[i+1..p]$ and any proper suffix
    $w[s+1..j]$ of $w[i+1..j]$
         we have $(G_w(p)-G_w(i), \bar{b}) >0$ and
     $(G_w(j)-G_w(s), \bar{b})<0$, which implies that $w[i+1..p]$ 
	and $w[s+1..j]$ do not have the same letter frequencies.
\end{proof}

\begin{proof}[Proof of Theorem~\ref{wap1}.]
    Like in the binary case, balance 
	implies that
    $G_w$  belongs to some $k$-dimensional cylinder $C (\emph{\textbf{l}},M) $.
     The coefficients $a_i$ of the axis $\emph{\textbf{l}}$ of the cylinder
    are related to the letter frequencies and can be defined via
    $\{{\bf{v}}_1, \dots, {\bf{v}}_{k}\}$.
    Since frequencies are rational and all ${\bf{v}}_i$ have integer entries, the
    coefficients $a_i$ are also rational.
    In fact, the graph of the word lies on finitely many lines with rational coefficients
    parallel to the cylinder axis.
    Clearly, in dimension two this is equivalent to bounded width.

    First notice that such tangential lines always exist.
    We can choose a cylinder with the minimal $M$.
    Similarly to the binary
    case and using the fact that the slope vector $\bar{a}$ of the
    cylinder is rational it is not hard to show that there exists a
    point $\bar{x}' \in G_w$ with $d(\emph{\textbf{l}}, \bar{x}') = M.$
    Consider the line $\emph{\textbf{l}}'$ parallel to
    $\emph{\textbf{l}}$ and passing through one of such points,
    $\emph{\textbf{l}}'\colon \bar{x} = \bar{x}' + \bar{a} t $.
    Taking the tangential plane $H$ to the cylinder passing through $\bar{x}'$,
    we see that $\emph{\textbf{l}}'$ is a tangential line to the graph
    $G_w$. We remark that this tangential plane $H$ is both a
    tangential plane to the graph $G_w$ and a ``proper'' tangential
    plane to the cylinder $C (\emph{\textbf{l}},M) $.
    Although there can be other tangential lines to the graph $G_w$ lying inside the
    cylinder, so that the corresponding tangential planes to $G_w$ are
    not tangential to the cylinder.

    To prove the theorem it remains to show that for a tangential line
    $\emph{\textbf{l}}''\colon \bar{x} = \bar{x}'' + \bar{a} t $ the points
    of the graph lie on this line with bounded gaps.

    Like in the binary case, we first show that there are infinitely
    many points of the graph on the line $\emph{\textbf{l}}''$.
    Suppose that there are only finitely many of those. Take the
    corresponding tangential hyperplane $H''\colon b_1 x_1 + \dots + b_k
    x_k = d''$ and consider the hyperplane $H'\colon b_1 x_1 + \dots + b_k
    x_k = d'$ parallel to $H''$ such that it has infinitely many
    points of $G_w$ on it (since $G_w$ lies on finitely many lines
    parallel to $\emph{\textbf{l}}''$ due to rational coefficients, it
    also lies on finitely many hyperplanes parallel to $H''$). So,
    there are finitely many points of $G_w$ such that $b_1 x_1 + \dots
    + b_k x_k < d'$, and infinitely many with $b_1 x_1 + \dots + b_k x_k = d'$.

    Now take the point $\bar{x}^{(n_0)}=G_w(n_0)$ on the line with
    the largest $n_0$ such that $b_1 x^{(n_0)}_1 + \dots + b_k x^{(n_0)}_k
    < d'$. Then for each $n$ with $b_1 x^{(n)}_1 + \dots + b_k x^{(n)}_k = d'$
    for $\bar{x}^{(n)}=G_w(n)$ the factor $w[n_0+1..n]$ is weakly abelian
    unbordered. Indeed, arguing as in Lemma \ref{wab_factor1}, for any
    proper prefix $w[i+1..p]$ and any proper suffix $w[s+1..j]$ of
    $w[i+1..j]$ we have $(G_w(p)-G_w(i), \bar{b}) >0$ and
     $(G_w(j)-G_w(s), \bar{b})<0$, which implies that $w[i+1..p]$ cannot
     be weakly abelian equivalent to $w[s+1..j]$.
    So, we proved that there are infinitely many points of the graph on a tangential line.

    Finally, we need to prove that the points on the tangential line
    $\emph{\textbf{l}}''$ appear in bounded gaps.
    It follows from the fact that for two consecutive points $n_1,$ $n_2$ on
    $\emph{\textbf{l}}''$ the factor $w[n_1+1..n_2]$ is weakly abelian
    unbordered by Lemma \ref{wab_factor1}.
\end{proof}

\section{Abelian borders}\label{ab}

In this section we consider the connections between abelian
periodicity and the condition of having finitely many abelian unbordered factors. 
First we show that finitely many abelian unbordered
factors is not a sufficient condition for periodicity:

\begin{proposition}\label{prop_abborders}
    There exists an infinite aperiodic word $w$ and constants $C, D$
    such that every factor $v$ of $w$ with $|v|\geq C$ has an abelian
    border of length at most $D$.
\end{proposition}

\begin{proof}
    Any aperiodic word $w \in \{ 0101 00110011, 0101 0011 0011 0011 \}^\omega$
    satisfies the condition with $C=15$ and $D=14$.
   The proof is a fairly straightforward and not too technical case study:

    If $u$ is a sufficiently long abelian unbordered factor, then it
    must begin or end with either $0100$ or with $1101$ because the other factors of
    length $4$ have equal frequencies.
    So there are only these two cases to consider: the factor
    starts with $0100$ (the case of ending with $1101$ is symmetric) or
    with $1101$ (the case of ending with $0100$ is symmetric).

    Consider first the case of a factor beginning with $0100$.
    If it ends with $0$, we have an abelian border of length $1$;
    if it ends with $01$, we have an abelian border of length $2$.
    So we need to analyze the remaining case of the form $0100 \cdots 0011$.

    Considering extension of the prefix $0100$ to the right, one has
    after $01$ abelian period $4$ with equal frequencies.
    Considering extension of the suffix $0011$ to the left, one has abelian period
    $4$ with equal frequencies.
    So, extending to the left the suffix, one will get an abelian border as soon as there is a block $0101$.
    Thus, there exists an abelian border of length $4k+2$, where $k$ is at most 3.

    Now consider the case of a factor beginning with $1101$. 
	It continues uniquely to the right with $01001100$, so that $110101001100$ is the prefix of the factor.
    If it ends with $1$, we have an abelian border of length $1$,
    so we need to consider the cases of ending with $00$ and $10$.

    If it ends with $00$, then it has either a suffix $0100$, or
    $1100$. The suffix $0100$ extends uniquely to the left to
    $11010100$, giving an abelian border of length $8$. The
    suffix $1100$ extends to either $11001100$ giving an abelian border of length
    $8$, or to $110101001100$, giving an abelian border of length $12$.

    If it ends with $10$, then it has a suffix $010$ (otherwise there is an abelian border of length $3$), which
    extends uniquely to $1010$. This suffix in turn extends to
    $01010$ (otherwise there is an abelian border of length $5$),
    which extends to $1101010$, giving an abelian border of length
    $7$. This concludes the proof.
\end{proof}

We observe that all infinite word $w \in \{ 0101 00110011, 0101 0011 0011 0011 \}^\omega$ 
occurrring in the proof of  Proposition \ref{prop_abborders}
are abelian periodic with period $2$. 
In general, we do not know whether an infinite word containing 
only finitely many abelian unbordered factors is necessarily abelian periodic.  
However, we begin by showing that abelian periodicity alone 
does not in general imply only finitely many abelian unbordered factors. 
In other words, at least one direction of Theorem~1 fails in the abelian setting.

\begin{proposition}\label{pal} 
	Let $x$ be a uniformly recurrent aperiodic word containing infinitely many distinct palindromes. 
	Then $x$ admits an infinite number of abelian unbordered factors.
\end{proposition}

\begin{proof} 
	Let $N$ be any positive integer. 
	We claim that $x$ admits a factor of the form $aUb$ where $a$ and $b$ are distinct letters 
	and $U$ a palindrome with $|U|\geq N$. 
	In fact, since $x$ contains infinitely many palindromes, 
	there exists a palindromic factor $u$ of $x$ with $|u|\in \{N,N+1\}$. 
	By uniform recurrence there exists a constant $M$ such that every factor of $x$ 
	of length $M$ contains an occurrence of $u$. 
	Since $x$ is aperiodic, there exists a word $v$ with $|v|=3M$ and distinct letters $c$ and $d$ 
	such that both $cv$ and $dv$ are factors of $x$. 
	In the usual terminology, $v$ is a left special factor of $x$. 
	Let $\eta$ denote the center of the first occurrence of $u$ in the prefix of $v$ of length $M$. 
	So $\eta$ is either an integer or a half integer depending on the parity of $|u|$. 
	Let $U$ be the longest palindromic factor of $v$ centered at $\eta$. 
	It follows that $2M\geq|U|\geq |u| \geq N$. 
	If $U$ is a not a prefix of $v$, then there exist distinct letters $a$ and $b$ 
	such that the word $aUb$ is a factor of $v$ centered at $\eta$. 
	On the other hand, if $U$ is a prefix of $v,$ then let $b$ be a letter such that $Ub$ is a prefix of $v$. 
	Since both $cv$ and $dv$ are factors of $x$ we can find a letter $a$ different from $b$ such that $aUb$ is a factor of $x$. 
	Having established our claim, the result of the proposition now follows 
	immediately since any word of the form $aUb$ with $a\neq b$ and $U$ a palindrome is clearly abelian unbordered.  
\end{proof}

As an immediate corollary we have:

\begin{corollary}\label{cor2}
	There exist abelian periodic words having infinitely many abelian unbordered factors.
\end{corollary}

\begin{proof} 
Let $x\in\{0,1\}^\omega$ be any binary uniformly recurrent aperiodic word 
containing infinitely many palindromes.  
For instance $x$ can be taken to be the Thue-Morse infinite word or any Sturmian word. 
Let $\tau$ denote the morphism $0\mapsto 0110$, $ 1\mapsto 1001$.  
Then $\tau(x)$ is abelian periodic (with abelian period $2$). 
Moreover, it is readily verified that $\tau(x)$ is uniformly recurrent, 
aperiodic, and contains infinitely many distinct palindromes.  
By Proposition~\ref{pal}, it follows that $\tau(x)$  
admits infinitely many abelian unbordered factors. 
\end{proof}

As mentioned earlier, we do not know whether the other direction 
of Theorem~\ref{Theorem1} 
holds in the abelian context, 
that is whether a word having only finitely many abelian unbordered factors 
is necessarily abelian periodic. 
However we are able to establish the abelian analogue of the following 
weaker version of Theorem~\ref{Theorem1}:

\begin{theorem}\label{comp}
	Let $x$ be an infinite word having only finitely many unbordered factors. 
	Then there exists a constant $N$ such that $x$ contains at most $N$ factors 
	of each given length  $n\geq 1$. 
	In other words, $x$ has bounded factor complexity.
\end{theorem}

The reason this is a weaker version of Theorem~\ref{Theorem1} 
is because bounded complexity is equivalent to ultimate periodicity. 
For this reason the converse of Theorem~\ref{comp} does not hold in general. 
For instance $x=01111\cdots$ has bounded complexity, 
yet contains infinitely many unbordered prefixes.

We now give several new proofs of Theorem~\ref{Theorem1} 
which may be of independent interest. 
One direction is of course trivial: 
namely, if x is purely periodic then all sufficiently long factors of x are bordered. 
For the other direction,  as in the original proof by  Ehrenfeucht and Silberger,  
we break the proof into two parts; 
in a first part we show that if x is any infinite word with the property 
that all sufficiently long factors are bordered, then x is ultimately periodic. 
We give three proofs of this first part which to our knowledge are all new. 
However, each of the following proofs relies on one 
or more key features which do not extend to the abelian context. 
See the remark following the proofs.
In a second part we show that x must be purely periodic. 
Also our proof of the  second part is new and greatly simplifies 
the original proof by Ehrenfeucht and Silberger.

\begin{proof}[Proof of  Theorem~\ref{Theorem1}]
Let  $\mathcal{U}(x)$ denote the set of unbordered factors of $x$. 
Clearly if $x$ is purely periodic then $\mathcal{U}(x)$ is finite. 
Conversely, assume  $\mathcal{U}(x)$ is finite. 
We first show that $x$ is ultimately periodic.  

A first quantitative proof of this was given by K. Saari  in which 
he shows that for any aperiodic word $x$ and any positive integer $n$,   
the number $\mathcal{L}_x(n)$ of Lyndon words of length $\leq n$ occurring in $x$ 
is greater or equal to $\mathcal{L}_{{\mathbf f}}(n)$ where $\mathbf f$ is the Fibonacci word. 
See  Theorem~3 and Remark~3 in \cite{Saari}. 

A second proof, also using Lyndon words proceeds as follows: 
If $x$ is aperiodic, then by Proposition~4.8 in \cite{SS}, 
$x$ contains an infinite Lyndon word $x'$ in its shift orbit closure, 
i.e.,  there exists an ordering of the alphabet of $x$ relative to which 
$x'$ is lexicographically smaller than all its proper suffixes. 
Since $x'$ begins in infinitely many unbordered prefixes, 
it follows that $x$ contains infinitely many unbordered factors. 

A third proof uses the so-called Duval conjecture, 
first proved by the second author together with D. Nowotka in \cite{HN}, 
which states that if $v$ is a word of length $2n$ 
such that the prefix $u$ of $v$ of length $n$ is unbordered 
and every factor of $v$ of length greater than $n$ is bordered, then $v=uu$. 
Assume  $\mathcal{U}(x)$ is finite and let $u\in  \mathcal{U}(x)$ be of maximal length. 
Then by the Duval conjecture, every occurrence of $u$ in $x$ is an occurrence of $u^2$. 
Hence $x$ is ultimately periodic with period $u$. 

A fourth proof, perhaps the simplest in that it is completely self contained, 
proceeds as follows: Assume $\mathcal{U}(x)$ is finite. 
By replacing $x$ if necessary by some suffix of $x$, 
we can assume that each $u\in \mathcal{U}(x)$ is recurrent in $x$, 
i.e., occurs in $x$ an infinite number of times. 
Let $u\in \mathcal{U}(x)$ be of maximal length. 
Let $v$ be any first return to $u$ in $x$, i.e., 
$vu$ is a factor of $x$ which begins and ends in $u$ and $|vu|_u=2$. 
Then since $u$ is unbordered, $|v|\geq|u|$ i.e., $u$ is a prefix of $v$.  
If $|v|>|u|$, then $v$ is bordered (by maximality of $|u|)$. 
Let $v'$ denote the shortest border of $v$. 
Then $v'\in \mathcal{U}(x)$ from which it follows that $|v'|\leq|u|$ 
and hence $v'$ is a proper prefix of $u$ (since $u$ is not a suffix of $v)$.  
But then the factor $v'u$ of $x$ is an unbordered factor of $x$ 
of length greater than $|u|$, a contradiction. 
Hence, if $v$ is a first return to $u$, then $|v|=|u|$, i.e., $v=u$. 
It follows that $x$ is ultimately periodic with period $u$. 

We next show that if $x=x_1x_2x_3\cdots $ is ultimately periodic 
and all sufficiently long factors of $x$ are bordered, then $x$ is  purely periodic. 
Let $x'$ denote the left infinite word $\cdots x_3x_2x_1$ obtained by reflecting $x$. 
By assumption there exists a primitive word $u$
such that  $\prescript{\omega\mkern -2mu}{}u$ 
is a prefix of $x'$ 
(where $\prescript{\omega\mkern -2mu}{}u$ 
denotes the left infinite word $\cdots uuu$). 
Clearly if $u=a$ for some letter $a$, then $x'=\prescript{\omega\mkern -2mu}{}a$. 
Thus we can assume that $u$ contains at least two distinct letters. 
Let $<$ be a linear order of $\alph(x)$ and let $<'$ denote the opposite order, 
i.e., $a<'b$ if and only if $b<a$. 
 Let $M$ ($m$, respectively) denote the Lyndon conjugate of $u$ 
relative to $<$ ($<'$, respectively). 
Without loss of generality we can write 
$x'=\prescript{\omega\mkern -4mu}{}Mz=\prescript{\omega\mkern -2mu}{}myz$ 
for some suffix $z$ of $x'$ and some suffix $y$ of $M$. 
We will show that $z$ is a prefix of $M^\omega$ 
and hence that $x'$ is  purely periodic.
Let $M_z$ denote the prefix of $M^{\omega}$ of length $|z|$ 
and $m_{yz}$ the prefix of $m^{\omega}$ of length $|yz|$. 
Since $M$ is unbordered and all sufficiently long factors of $x'$ are bordered, 
it follows that $z$ is a product of prefixes of $M$ and hence that $z\leq M_z$. 
Similarly, $yz\leq' m_{yz}$ or equivalently $yz\geq m_{yz}$. 
 Since $m^\omega =yM^\omega$, we deduce that $z\geq M_z$ 
and hence $z=M_z$ as required.
\end{proof} 

\begin{remark}
We note that none of the above proofs readily extend to the abelian context. 
In fact, we know of no suitable analogue of Lyndon words in the abelian setting. 
Also,  the Duval conjecture fails miserably in the abelian setting. 
More precisely, there exists an infinite word $x$ with the following properties: 
i) $x$ begins in an unbordered factor $u,$ 
ii) all factors of $x$ of length greater than $|u|$ are abelian bordered, 
iii) the prefix of $x$ of length $2|u|$ is not an abelian square. 
To see this, let $y$ be any infinite word in $\{ 0101 00110011, 0101 0011 0011 0011 \}^\omega$ 
beginning in $(0101 0011 0011 0011 )^2$.  
Let $x$ denote the second shift of $y,$ i.e., $x=(01)^{-1}y$. 
Then $x$ begins in the unbordered factor $u=01001100110011$ of length $14$. 
As in the proof of Proposition~\ref{prop_abborders}, 
every factor of $x$ of length greater than $14$ is abelian bordered. 
Yet it is easily checked that the prefix of length $28$ of $x$ is not an abelian square, i.e., 
not of the form $uu'$ with $u'$ abelian equivalent to $u$. 
Finally, the fourth proof uses the following fact which is no longer true in the abelian setting, 
namely that  if $v$ is a proper prefix of an unbordered word $u,$ then $vu$ is unbordered.  
For instance, we can take the prefix $v=01$ of the abelian unbordered word $u=010011$. 
Yet $vu=01010011$ is abelian bordered. 
\end{remark}





We end by establishing the following abelian analogue of Theorem~\ref{comp}:

\begin{theorem}Let $x$ be an infinite word having only finitely many abelian unbordered factors. 
Then there exists a constant $N$ such that $x$ contains 
at most $N$ abelian equivalence classes of factors of each given length  $n\geq 1$. 
In other words, $x$ has bounded abelian factor complexity.
\end{theorem}

\begin{proof} Since the condition that $x$ has only finitely many
abelian unbordered factors implies that $x$ has only  finitely many
weakly abelian unbordered factors, it follows that all
conclusions we derived earlier in the weak abelian setting still hold. 
In particular $x$ is balanced. Thus $x$ has bounded abelian complexity (see Lemma~3 in \cite{RSZ}).
\end{proof}

\section{A non-abelian periodic word with bounded abelian square at each position}\label{5}

In \cite{akp}, the following open question was proposed: Let $w$
be an infinite word and $C$ be an integer such that each position
in $w$ is a centre of an abelian square of length at most $C$. Is
$w$ abelian periodic?  We answer this question negatively by
providing an example (actually, a family of examples).

Consider a family of infinite words of the following form:
$$( 000101010111000111000 (111000)^* 111010101 )^{\omega}$$
A straightforward case study shows that words of this form have an
abelian square of length at most $12$ at each position. It is not
hard to see that this family contains abelian aperiodic words.

\section{Conclusions and open questions}

We conclude with the summary of our results and propose two open problems.

\begin{question}\label{question1}
Let $w$ be an infinite word and $C$ a constant such that every
factor $v$ of $w$ with $|v|\geq C$ is abelian bordered. Does it
follow that $w$ is abelian periodic?
\end{question}

Recall that we showed that there exist examples of words
satisfying these conditions which are not periodic (Proposition
\ref{prop_abborders}), but all examples we have are abelian
periodic.

\medskip

The next question asks whether the converse of Theorem \ref{wap1}
(and in particular Theorem \ref{wap}) is true:

\begin{question}  \label{question2}
   Let $w$ be an infinite bounded weakly abelian periodic word over a $k$-letter alphabet
  such that its graph $G_w$ belongs to a $k$-dimensional cylinder with axis with rational coefficients,
    and each tangential line to $G_w$ has points of $G_w$ on it with bounded
    gaps. Does it follow that $w$ has only finitely many weakly
   abelian unbordered factors?
\end{question}

We remark that in the case of binary alphabets and letter frequencies equal to $1/2$
the answer to this question is positive
(see Remark after Theorem  \ref{wap}).

Our main results and open questions are summarized in Figure 1.

\small

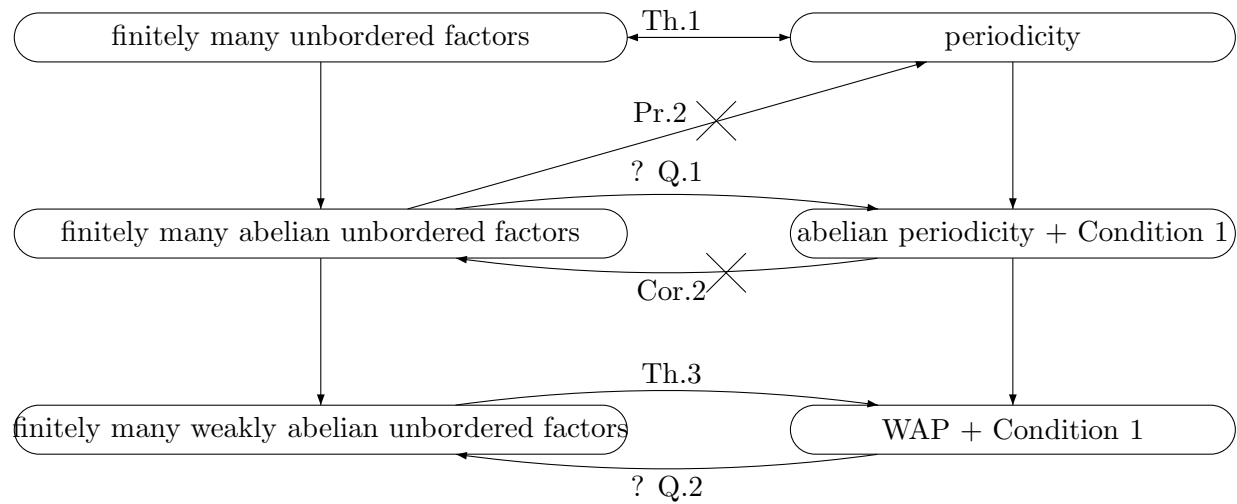
\begin{figure}[h]
      \begin{center}
        \unitlength=3.7pt
            \begin{picture}(50, 50)(0,-5)
              \gasset{Nw=62,Nh=5,Nmr=2.5,curvedepth=0}
              \thinlines
              \put(5,0){ \node(A1)(-10,40){finitely many unbordered factors}
              \gasset{Nw=45,Nh=5,Nmr=2.5,curvedepth=0}
              \node(A2)(60,40){periodicity}

              \drawedge(A1,A2){ Th.\ref{Theorem1}}
              \drawedge(A2,A1){}

        \gasset{Nw=62,Nh=5,Nmr=2.5,curvedepth=0}
                    \node(A3)(-10,20){finitely many abelian unbordered factors}
              \gasset{Nw=45,Nh=5,Nmr=2.5,curvedepth=4}
              \node(A4)(60,20){abelian periodicity $+$ Condition 1}

              \drawedge(A3,A4){? Q.\ref{question1}}
              \drawedge(A4,A3){ Cor.\ref{cor2}}
              \drawline[AHnb=0](29,14)(33,18)
              \drawline[AHnb=0](33,14)(29,18)
              \gasset{Nw=62,Nh=5,Nmr=2.5,curvedepth=0}
                \drawedge(A3,A2){Pr.\ref{prop_abborders}}
              \drawline[AHnb=0](28,29.5)(32,33.5)
              \drawline[AHnb=0](32,29.5)(28,33.5)
                 \drawedge(A1,A3){}
                  \drawedge(A2,A4){}

                    \node(A5)(-10,0){finitely many weakly abelian unbordered factors}
              \gasset{Nw=45,Nh=5,Nmr=2.5,curvedepth=0}
              \node(A6)(60,0){WAP $+$ Condition 1}

        \gasset{Nw=45,Nh=5,Nmr=2.5,curvedepth=4}
              \drawedge(A5,A6){ Th.\ref{wap1}}
              \drawedge(A6,A5){? Q.\ref{question2}}
              \gasset{Nw=45,Nh=5,Nmr=2.5,curvedepth=0}
                 \drawedge(A3,A5){}
                  \drawedge(A4,A6){}
                  }

            \end{picture}
    \end{center}
         \caption{Results and open questions.
     WAP means  ``weakly abelian periodic".
    Condition 1: The graph $G_w$ of the word belongs
         to a $k$-dimensional cylinder with axis with rational coefficients,
    and each tangential line to $G_w$ has points of $G_w$ on it with bounded
    gaps.}
        \label{table}
\end{figure}

\end{document}